\documentclass[11pt,a4paper]{article}

\usepackage{a4}
\usepackage{amsmath}
\usepackage{amssymb}
\usepackage{axodraw2}
\usepackage{bbm}
\usepackage{color}
\usepackage{graphicx}
\usepackage{jheppub}
\usepackage{latexsym}
\usepackage{mathtools}
\usepackage{pifont}
\usepackage{slashed}

\newcommand{\Neps}{n_\epsilon}
\newcommand{\alphas}{\alpha_s}

\newcommand{\cA}{{\cal A}}

\def\eq{{Q_{q}}}
\def\eqSq{{Q_{q}^{2}}}

\def\CDR{{\scshape cdr}}
\def\DRED{{\scshape dred}}

\def\FDH{{\scshape fdh}}
\def\FDR{{\scshape fdr}}

\def\HV{{\text{\scshape{hv}}}}

\def\CF{{C_F}}

\def\MS{{\overline{\text{\scshape{ms}}}}}

\def\dim{{d}}
\def\mDRED{{\rm\scriptscriptstyle DRED}}
\def\mCDR{{\rm\scriptscriptstyle CDR}}
\def\mHV{{\rm \scriptscriptstyle HV}}
\def\mFDH{{\rm \scriptscriptstyle FDH}}

\allowdisplaybreaks

\begin{document}
\thispagestyle{empty}

\begin{flushright}
PSI-PR-19-27\\
ZU-TH 55/19\\
\end{flushright}
\vspace{3em}
\begin{center}
{\Large\bf Dimensional schemes for cross sections at NNLO}
\\
\vspace{3em}
{\sc
C.\,Gnendiger$^{a,}$\footnote{e-mail: Christoph.Gnendiger@psi.ch},
A.\,Signer$^{a,b}$
}\\[2em]
{\sl ${}^a$ Paul Scherrer Institut,\\
CH-5232 Villigen PSI, Switzerland \\
\vspace{0.3cm}
${}^b$ Physik-Institut, Universit\"at Z\"urich, \\
Winterthurerstrasse 190,
CH-8057 Z\"urich, Switzerland}
\setcounter{footnote}{0}
\end{center}
\vspace{6ex}

\begin{center}
\begin{minipage}{15.3truecm}
{So far, the use of different variants of dimensional regularization
  has been investigated extensively for two-loop virtual
  corrections. We extend these studies to real corrections that are
  also required for a complete computation of physical cross sections
  at next-to-next-to-leading order. As a case study we consider
  two-jet production in electron-positron annihilation and describe
  how to compute the various parts separately in different schemes. In
  particular, we verify that using dimensional reduction the
  double-real corrections are obtained simply by integrating the
  four-dimensional matrix element over the phase space. In addition,
  we confirm that the cross section is regularization-scheme
  independent.}
\end{minipage}
\end{center}

%\vspace{0.5cm}
%\centerline
%{\small PACS numbers: 11.10.Gh, 11.15.-q, 12.38.Bx}

\newpage

 \noindent\hrulefill
  \tableofcontents
 \noindent\hrulefill

\bigskip
\section{Introduction}
\label{sec:introduction}

Beyond leading order, physical cross sections are usually computed as
sums of several terms that are individually divergent. These divergences
stem from the ultraviolet (UV) and the infrared (IR) regions of momentum
integrals. In most applications such divergences are dealt with by using
dimensional regularization, i.\,e.\ by working in $d\!=\!4\!-\!2\epsilon$
dimensions. This renders intermediate expressions well-defined with
divergences manifest as $1/\epsilon^n$ poles. While it is mandatory to
treat integration momenta in $d$ dimensions, there is quite some
freedom on how to treat other quantities in such computations. Hence,
in practice there is a variety of dimensional schemes that can be
used, the most common being conventional dimensional regularization
(\CDR), the 't~Hooft-Veltman scheme (\HV)~\cite{'tHooft:1972fi}, the
four-dimensional helicity scheme (\FDH)~\cite{BernZviKosower:1992},
and dimensional reduction (\DRED)~\cite{Siegel:1979wq}. For an
overview and a discussion of the basic properties of these schemes we
refer to~\cite{Gnendiger:2017pys} and references therein.

As it might be advantageous to use different regularization schemes for
different parts of the calculation, the relations between the various
schemes have to be understood. Starting this program, transition rules
for UV-renormalized virtual amplitudes at next-to-leading order (NLO)
have first been worked out in~\cite{Kunszt:1994} for massless QCD and
were then generalized to the massive case~\cite{Catani:2000ef}.
The regularization-scheme independence of cross sections at NLO is discussed
in~\cite{Catani:1996pk} and a recipe on how to compute consistently the
various ingredients (virtual, real, and initial-state collinear counterterm)
for hadronic collisions is given in~\cite{Signer:2005,Signer:2008va}.
The key observation is that so-called $\epsilon$-scalars have to be
introduced and to be considered as independent from $d$-dimensional
gluons. Going beyond NLO, a lot of work has been done to understand the UV
renormalization~\cite{Capper:1979ns,Jack:1993ws,Jack:1994bn,Harlander:2006rj,
Harlander:2006xq} as well as the virtual two-loop contributions~%
\cite{Kilgore:2011ta,Kilgore:2012tb,Gnendiger:2014nxa,Broggio:2015ata,
Broggio:2015dga, Gnendiger:2016cpg} in schemes other than \CDR.
The current status can be summarized as follows: for all dimensional schemes
mentioned above, UV renormalization and virtual corrections are understood at
least up to next-to-next-to leading order (NNLO), while the computation of
real corrections is only fully understood at NLO. A first step towards NNLO
for the latter has been made in~\cite{Czakon:2014oma} where the \HV~scheme
is used for real corrections. It is the purpose of the present paper to make
further progress on the scheme dependence of real contributions and to
discuss the calculation of double-real and real-virtual corrections at
NNLO in different schemes. In particular, we will focus on \DRED\ as
this is the most general dimensional scheme usually considered.

To this end, we consider NNLO QCD corrections to the process
$e^{+}e^{-}\!\to\!2~jets$ and compute the double-virtual, double-real,
and real-virtual contributions separately in \DRED. This is an
extension of the corresponding computation in
\CDR~\cite{Anastasiou:2004qd, GehrmannDeRidder:2004tv}.  Moreover, we
perform the computations also in \HV\ and \FDH\ and show that the
physical cross section is regularization-scheme independent.
The process at hand has also been considered in non-dimensional regularization
schemes, i.\,e.\ schemes that keep the integration momenta in strictly four
dimensions. In fact, the fermionic contributions have been computed recently~%
\cite{Page:2018ljf} using 'four-dimensional regularization' (\FDR)~%
\cite{Pittau:2012zd, Donati:2013voa}. Similar to 'implicit regularization'~%
\cite{Sampaio:2005pc, Fargnoli:2010ec, Cherchiglia:2010yd} and 'loop
regularization'~\cite{Wu:2003dd, Bai:2017zuw}, \FDR\ is a four-dimensional
framework to compute higher-order corrections. Another approach that is being
investigated is to use loop-tree duality to deal with IR singularities at the
integrand level~\cite{Hernandez-Pinto:2015ysa, Driencourt-Mangin:2019aix}.
While these are interesting developments, they typically require that the full
computation is performed in the corresponding scheme. It will be very difficult
to combine partial results obtained in dimensional schemes with computations in
non-dimensional schemes. Hence, in this work we focus on the former.

We start in Section~\ref{sec:prelim} with a brief recapitulation of the
most important aspects of the dimensional schemes before we consider QCD
corrections to the process $e^{+}e^{-}\!\to\!\gamma^{*}\to\!q\bar{q}$
in Section~\ref{sec:process}. This section contains the main results of
the paper including a description on how to compute double-virtual,
double-real, and real-virtual corrections in \DRED. We also show that the
total cross section is scheme independent, as required.
The particular role of the $\epsilon$-scalars is investigated in
Section~\ref{sec:epsContributions}. The computation of the cross section
in \FDH\ is discussed in Section~\ref{sec:fdh} together with the \HV~scheme,
before we conclude in Section~\ref{sec:conclusions}.\\

\section{Dimensional schemes}
\label{sec:prelim}

As mentioned before, an efficient way to regularize UV and IR divergences
at the same time is to formally shift the dimension of loop and phase-space
integrations from (strictly) four to
\begin{align}
 \dim\equiv4-2\,\epsilon\,,
\end{align}
with an arbitrary regularization parameter $\epsilon$. In this way, divergent
integrals are parametrized in terms of $1/\epsilon^n$ poles. Although not
strictly necessary, it is usually advantageous to also modify the dimensionality
of other algebraic objects. The most commonly used approach in this respect
is \CDR, where \textit{all} Lorentz indices are considered in quasi $\dim$
dimensions. Indicating the dimension by a subscript, we therefore write
\begin{align}
  \text{\CDR}:\qquad
  k^{\mu}_{[\dim]},\,
  \gamma^{\mu}_{[\dim]},\,
  g^{\mu\nu}_{[\dim]},\,
  A^{\mu}_{[\dim]},\,
  \dots\,
  \phantom{\Big|}
\end{align}
for loop momenta, $\gamma$ matrices, metric tensors, vector fields, etc.
From a conceptual point of view this approach is the simplest realization of
dimensional regularization in the sense that all dimensionful quantities are
treated on the same footing. As a consequence, it is for example sufficient
to impose one single (modified) Lorentz algebra. However, it is important to
realize that this formal simplicity does not automatically guarantee that
\CDR\ is also the best choice regarding computational efficiency.

A second realization of dimensional regularization is \DRED\ where all
dimensionful quantities except for loop momenta are treated
in quasi $\dim_s$ dimensions with
\begin{align}
 \dim_s\equiv\dim+\Neps\,.
 \label{eq:ds}
\end{align}
The value of $\dim_s$ does not necessarily have to be fixed as long as the
limit $\Neps\!\to\!0$ is implied at the end of the calculation. Usually,
however, it is taken to be $d_s\!=\!4$, resulting in $\Neps\!=\!2\epsilon$.
The dimensionful quantities from above are accordingly written as
\begin{align}
  \text{\DRED}:\qquad
  k^{\mu}_{[\dim]},\,
  \gamma^{\mu}_{[\dim_s]},\,
  g^{\mu\nu}_{[\dim_s]},\,
  A^{\mu}_{[\dim_s]},\,
  \dots\,\,.
  \phantom{\Big|}
\end{align}
One important aspect of \DRED\ is that (in contrast to what the name of
the scheme suggests) the underlying vector space is 'bigger' than the one
of \CDR, as indicated by \eqref{eq:ds}. Thus, in spite of $d_s\!=\!4$, for
consistency the vector space of \DRED\ is in fact infinite-dimensional~%
\cite{ Stockinger:2005gx}. Therefore, it is always possible to split quasi
$\dim_s$-dimensional quantities into a quasi $\dim$-dimensional '\CDR\ part'
and an evanescent remainder, e.\,g.
\begin{align}
 A^{\mu}_{[\dim_s]}\!=\!A^{\mu}_{[\dim]}\!+\!A^{\mu}_{[\Neps]}\,
 \phantom{\Big|}
  \label{eq:vectorFieldSplit}
\end{align}
for vector fields.
The field $A^{\mu}_{[\Neps]}$ is often referred to as $\epsilon$-scalar.
In the case of \FDH\ and \HV\ we also need strictly four-dimensional
quantities such as $\gamma^{\mu}_{[4]}$, $g^{\mu\nu}_{[4]}$, and
$A^{\mu}_{[4]}$, as discussed in~\cite{Gnendiger:2017pys}.

The application of the schemes mentioned above to the computation of
two-loop virtual corrections is well understood and leads to universal
scheme dependences which can be described in terms of
scheme-dependent IR anomalous dimensions. Since physical cross sections
must not depend on the regularization scheme, these dependences have
to be canceled once real and real-virtual corrections are added.
At NLO, this has been studied in detail~\cite{Signer:2008va}. It was
found that \CDR\ and \DRED\ are unitary in the sense that this cancellation
is ensured by simply integrating the corresponding squared matrix
elements over the phase space. This is related to the fact that in
these two schemes 'regular' and 'singular' vector bosons are treated
equally.%
\footnote{ See \cite{Gnendiger:2017pys} for more details.  In previous
  papers like~\cite{Signer:2008va}, the somewhat misleading terms
  'external' and 'internal' have been used for 'regular' and
  'singular', respectively.}  In particular, in \DRED\ the real
corrections are obtained by evaluating the matrix element in
$\dim_s\!=\!4$ dimensions and by integrating it over the
$\dim$-dimensional phase space.
In \HV\ and \FDH, regular and singular vector bosons are treated
differently and these schemes are not unitary. It is still possible to
use these schemes consistently, but the $\mathcal{O}(\epsilon)$ terms
of the real matrix elements that arise from singular regions have to
be taken into account properly.

The argument that in \DRED\ the real corrections are consistently
obtained by integrating the corresponding four-dimensional matrix
element over the $\dim$-dimensional phase space is independent of
the order of perturbation theory.  One main objective of the present
paper is to show that \DRED\ is indeed a consistent and unitary
regularization at NNLO also for the IR regions.  In order to do so,
we (re)derive the well-known analytic NNLO result~%
\cite{Celmaster:1979xr,Chetyrkin:1979bj} of the QCD corrections to
$e^{+}e^{-}\!\to\gamma^{*}\!\to\!2~jets$ using \CDR\ and \DRED,
respectively, and show that the partonic cross section is a
regularization-scheme independent quantity. We do not use the optical
theorem but compute the virtual, real, and real-virtual contributions
separately in both schemes, as done for \CDR\ in
\cite{GehrmannDeRidder:2004tv}. To our knowledge, this is the first
time that \DRED\ is used for real (and real-virtual) corrections at
NNLO.\\

\section{The process $e^{+}e^{-}\!\to\gamma^{*}\!\to\!q\bar{q}$ in DRED}
\label{sec:process}
\subsection{Tree-level contribution}

 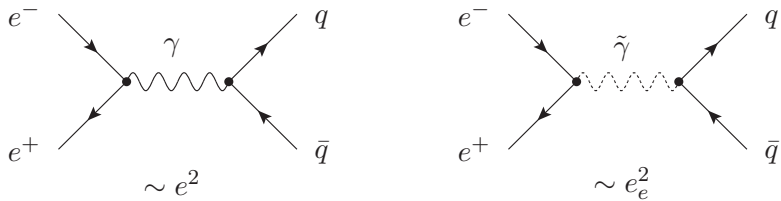
\begin{figure}[t]
 \begin{center}
 \scalebox{.85}{
 \begin{picture}(135,90)(0,0)
  \Vertex(45,45){2}
  \Vertex(90,45){2}
  \ArrowLine(45,45)(15,15)
  \ArrowLine(15,75)(45,45)
  \ArrowLine(120,15)(90,45)
  \ArrowLine(90,45)(120,75)
  \Photon(45,45)(90,45){4}{4}
  \Text(0,75){\scalebox{1.18}{$e^{-}$}}
  \Text(0,15){\scalebox{1.18}{$e^{+}$}}
  \Text(135,75){\scalebox{1.18}{$q^{\phantom{-}}$}}
  \Text(135,15){\scalebox{1.18}{$\bar{q}^{\phantom{+}}$}}
  \Text(65,60){\scalebox{1.18}{$\gamma$}}
  \Text(65,0){\scalebox{1.18}{$\sim e^2$}}
  \end{picture}
  }
  \qquad\qquad
  \scalebox{.85}{
  \begin{picture}(135,90)(0,0)
  \Vertex(45,45){2}
  \Vertex(90,45){2}
  \ArrowLine(45,45)(15,15)
  \ArrowLine(15,75)(45,45)
  \ArrowLine(120,15)(90,45)
  \ArrowLine(90,45)(120,75)
  \DashPhoton(45,45)(90,45){4}{4}{1.5}
  \Text(0,75){\scalebox{1.18}{$e^{-}$}}
  \Text(0,15){\scalebox{1.18}{$e^{+}$}}
  \Text(135,75){\scalebox{1.18}{$q^{\phantom{-}}$}}
  \Text(135,15){\scalebox{1.18}{$\bar{q}^{\phantom{+}}$}}
  \Text(65,60){\scalebox{1.18}{$\tilde{\gamma}$}}
  \Text(65,0){\scalebox{1.18}{$\sim e_e^2$}}
  \end{picture}
  }
  \end{center}
  \caption{Diagrams contributing to the process
   $e^{+}e^{-}\!\to\gamma^{*}\!\to\!q\bar{q}$ at the tree level.
    The interaction is mediated by a quasi $\dim$-dimensional
    photon~$\gamma$ (left) and a quasi $\Neps$-dimensional $\epsilon$-scalar
    photon~$\tilde{\gamma}$ (right), respectively.
    The right diagram only exists in \DRED.}
  \label{fig:FdiagTree}
  \end{figure}
 
As a benchmark process to compare the characteristics of \CDR\ and
\DRED\ in the multi-loop regime we consider QCD corrections to the
process $e^{+}e^{-}\!\to\gamma^{*}\!\to\!2~jets$ at NNLO accuracy.
More precisely, we retrace the well-known analytical calculation in
\CDR~\cite{GehrmannDeRidder:2004tv} and compare it with the
corresponding one in~\DRED.
To start with, we consider the (spin summed/averaged) squared
tree-level matrix elements
$M_{}^{(0)}\!=\!\langle\mathcal{A}_{}^{(0)}\,
|\,\mathcal{A}_{}^{(0)}\rangle$ in both schemes, i.\,e.\
\begin{subequations}
\label{eq:born}
  \begin{align}
      M_{\mCDR}^{(0)}\,\
      =\ & %\frac{\eqSq\,N_c}{3\,s}\,e^4\,(d-2)
      %\equiv
      \omega^{(0)} \, e^4 \, (d-2)\, ,
      \phantom{\Big|}
      \label{eq:born_cdr}
      \\*
      M_\mDRED^{(0)}
      =\ & M_\mDRED^{(0,\gamma)} +  M_\mDRED^{(0,\tilde{\gamma})}
      %= M_\mCDR^{(0)} +  M_\mDRED^{(0,\tilde{\gamma})}
      =\omega^{(0)}\, \Big[ e^4\, (d-2) + e^4_{e}\,\Neps\Big]\,.
      \phantom{\Big|}
      \label{eq:born_dred}
     %  \\
     %  M_{\mFDH}^{(0)}\,\
     %  =\ & M_{\mHV}^{(0)}\,
     %  =\  \omega^{(0)} \, e^4 \, (4-2)\,.
     %  \phantom{\Big|}
     %  \label{eq:born_fdh}
    \end{align}
  \end{subequations}
The quantity $\omega^{(0)}\!\equiv\eqSq\,N_c/(3\,s)$ contains the
electric charge $\eq$ of the quark, its colour number $N_c$, and the
c.\,o.\,m.\ energy $s$; the flux factor $1/(2\,s)$ is already
included.  As shown in Figure\,\ref{fig:FdiagTree}, in \CDR\ one diagram
contributes at the tree level. It contains a quasi $\dim$-dimensional
photon $\gamma$ and is proportional to the electromagnetic gauge
coupling $e$. Making use of the split in \eqref{eq:vectorFieldSplit}
gives rise to a second diagram which contains a quasi $\Neps$-dimensional
$\epsilon$-scalar photon $\tilde{\gamma}$ and which only contributes in \DRED.
Its evanescent coupling $e_e$ to fermions is not affected by gauge symmetry
and therefore has in genearal to be distinguished from the gauge coupling~$e$.
At the tree level, such a distinction is of course not strictly
necessary as the different renormalization of $e$ and $e_e$ is of higher order
in the perturbative expansion~\cite{Capper:1979ns}. It is done here for later
purposes.

Integrating \eqref{eq:born} over the two-body phase space, one gets for
the Born cross section
\begin{subequations}
\begin{align}
  \sigma^{(0)}_{\mCDR}\,
    & %=\frac{\Phi_2(\epsilon)}{8\pi}\,M_{\mCDR}^{(0)}\
    = \omega^{(0)}
      \Big(\frac{e^4}{4\pi}\Big)\,
      (1-\epsilon) \, \Phi_2(\epsilon)\,,
  \label{eq:sigma0CDR}
  \\%[8pt]
  \sigma^{(0,\gamma)}_{\mDRED}
    & %=\frac{\Phi_2(\epsilon)}{8\pi}\,M_{\mDRED}^{(0,\gamma)} 
    \equiv  \sigma^{(0)}_{\mCDR}\,,
    \phantom{\Big(\frac{e_e^4}{4\pi}\Big)}
    \label{eq:sigma0DRED1}
  \\
    \sigma^{(0,\tilde{\gamma})}_{\mDRED}
    & %=\frac{\Phi_2(\epsilon)}{8\pi}\, M_{\mDRED}^{(0,\tilde{\gamma})} 
    = \omega^{(0)}
      \Big(\frac{e_e^4}{4\pi}\Big)\,
      \Big(\frac{\Neps}{2}\Big)\,\Phi_2(\epsilon)\,,
      \label{eq:sigma0DRED2}
\end{align}
\end{subequations}
with
\vspace{-0pt}
\begin{align}
 \Phi_2(\epsilon)
   = \Big(\frac{4\pi}{s}\Big)^\epsilon
    \frac{\Gamma(1-\epsilon)}{\Gamma(2-2\epsilon)}
  =1+{\cal O}(\epsilon)\,.
\end{align}
By construction, the tree-level contribution from the quasi
$\dim$-dimensional photon is the same in \CDR\ and \DRED, \eqref{eq:sigma0DRED1}.
Moreover, setting $\dim_s\!=\!4$ (and therefore $\Neps\!=\!2\epsilon$),
the following relations hold in the physical limit:
\begin{align}
 \sigma^{(0)}_{\mDRED}\Big|_{\epsilon\to0}
 =\Big[\sigma^{(0,\gamma)}_{\mDRED}
 +\sigma^{(0,\tilde{\gamma})}_{\mDRED}\Big]_{\epsilon\to0}
 \equiv\sigma^{(0,\gamma)}_{\mDRED}\Big|_{\epsilon\to0}
 \equiv\sigma^{(0)}_{\mCDR}\Big|_{\epsilon\to0}
 \equiv\sigma^{(0)}
 =\omega^{(0)} \Big(\frac{e^4}{4\pi}\Big)\,.
 \phantom{\bigg|}
 \label{eq:bornRel}
\end{align}
Of course, the prediction for the physical observable is the same in
\CDR\ and \DRED. This is even true for arbitrary values of the
(renormalized) evanescent coupling $e_e$; the corresponding
contribution is proportional to $\Neps\!=\!2\epsilon$ and therefore
vanishes in the physical limit $\dim\to4$.  In what follows, we show on
the one hand that the vanishing of evanescent contributions like in
\eqref{eq:sigma0DRED2} takes place also at higher orders of
perturbation theory. On the other hand, we explicitly demonstrate how
they can nevertheless help to facilitate the determination of the real
contributions and therefore of the computation as a whole.\\

\subsection{Virtual corrections}
\label{sec:doubleVirtual}
  
The double-virtual contributions to $e^{+}e^{-}\!\to\gamma^{*}\to\!q\bar{q}$
in \CDR\ are known for a long time \mbox{\cite{Kramer:1986sg,Matsuura:1987wt,
Matsuura:1988sm}} and can be expressed in terms of the UV-renormalized
one- and two-loop quark form factors as%
\footnote{The structures in the brackets stem from the fact that we have to
  integrate the expressions
    \mbox{$M^{(1)}=
  \!\langle\mathcal{A}_{}^{(0)}\,|\,\mathcal{A}_{}^{(1)}\rangle+
  \!\langle\mathcal{A}_{}^{(1)}\,|\,\mathcal{A}_{}^{(0)}\rangle$
  },
  \mbox{$M^{(2)}=
  \!\langle\mathcal{A}_{}^{(0)}\,|\,\mathcal{A}_{}^{(2)}\rangle+
  \!\langle\mathcal{A}_{}^{(2)}\,|\,\mathcal{A}_{}^{(0)}\rangle+
  \!\langle\mathcal{A}_{}^{(1)}\,|\,\mathcal{A}_{}^{(1)}\rangle$
  } over the phase space.}
\begin{align}
 \sigma^{(v)}_{\mCDR}
 =\sigma^{(0)}_{\mCDR}
  \Big[
    2 \,F^{(1)}_{q,\text{\CDR}}
  \Big]\,,
  \qquad
 \sigma^{(vv)}_{\mCDR}
 =\sigma^{(0)}_{\mCDR}
  \Big[
    2 \,F^{(2)}_{q,\text{\CDR}}
    + \big(F^{(1)}_{q,\text{\CDR}}\big)^2
  \Big]\,.
\end{align}
We list the results below for the sake of completeness and to fix our
conventions.  Setting $\mu^2\!=\!s$ for the regularization scale and defining
the usual strong gauge coupling $\alpha_s\!=\!g_s^2/(4\pi)$, the virtual one-
and two-loop corrections read
\begin{subequations}
\begin{align}
 \sigma^{(v)}_{\mCDR}
 = \sigma^{(0)}_{\mCDR} & \Big(\frac{\alphas}{4\pi}\Big)^{\phantom{2}}\CF\,
   %\Phi_2(\epsilon)\,s^{-\epsilon}
   %c_\Gamma(\epsilon)\,
 \Big\{
   \!-\!\frac{4}{\epsilon^2}
   -\frac{6}{\epsilon}
   -16
   +\frac{7\pi^2}{3}
   +\mathcal{O}(\epsilon)
 \Big\}\,,
 \\
\sigma^{(vv)}_{\mCDR}
= \sigma^{(0)}_{\mCDR} & \Big(\frac{\alphas}{4\pi}\Big)^2\CF\,\times\,
\notag\\*
\Big\{
C_F & \Big[
  \frac{8}{\epsilon ^4}
  +\frac{24}{\epsilon ^3}
  +\frac{
    82
    \!-\!\frac{28 \pi ^2}{3}
    }{\epsilon ^2}
  +\frac{
    \frac{445}{2}
    \!-\!26 \pi ^2
    \!-\!\frac{184 \zeta (3)}{3}
    }{\epsilon }
  +\frac{2303}{4}
  \!-\!86 \pi ^2
  \!-\!172 \zeta (3)
  \!+\!\frac{137 \pi ^4}{45}
  \Big]
\notag\\
+ \,
C_A & \Big[
  \frac{11}{\epsilon ^3}
  +\frac{
    \frac{32}{9}
    +\frac{\pi ^2}{3}
    }{\epsilon ^2}
  -\frac{
    \frac{961}{54}
    \!+\!\frac{11 \pi ^2}{6}
    \!-\!26 \zeta (3)
    }{\epsilon }
  -\frac{51157}{324}
  +\frac{1061 \pi ^2}{54}
  +\frac{626 \zeta (3)}{9}
  -\frac{8 \pi ^4}{45}
  \Big]
\notag\\
+ \, N_F & \Big[
  \!-\!\frac{2}{\epsilon ^3}
  -\frac{8}{9\, \epsilon ^2}
  +\frac{
    \frac{65}{27}
    \!+\!\frac{\pi ^2}{3}
    }{\epsilon }
  +\frac{4085}{162}
  \!-\!\frac{91 \pi ^2}{27}
  \!+\!\frac{4 \zeta (3)}{9}
  \Big]
+\mathcal{O}(\epsilon)
\Big\}\,.
\end{align}
\end{subequations}
The tree-level cross section is given in \eqref{eq:sigma0CDR}.

In order to obtain the corresponding results in \DRED, in a first step we split
the amplitudes in a similar way as in \eqref{eq:born_dred} and distinguish
contributions from a quasi $\dim$-dimensional photon~$\gamma$
%(which are proportional to the gauge coupling~$e$)
and contributions from a quasi $\Neps$-dimensional $\epsilon$-scalar
photon~$\tilde{\gamma}$,
%(which are proportional to the evanescent coupling~$e_e$)
 \begin{align}
   \sigma^{(v)}_{\mDRED}
    &=\sigma^{(v,\gamma)}_{\mDRED}
    +\sigma^{(v,\tilde{\gamma})}_{\mDRED}\,,
   %\\
   \qquad
   \sigma^{(vv)}_{\mDRED}
    =\sigma^{(vv,\gamma)}_{\mDRED}
    +\sigma^{(vv,\tilde{\gamma})}_{\mDRED}
   \,.
  \label{eq:split}
 \end{align}
%\begin{align}
%  \sigma^{(\text{virt.})}_{\mDRED}%(e,e_e)
%  & \equiv\sigma^{(\text{virt.},\gamma)}_{\mDRED}%(e)\,
%   +\sigma^{(\text{virt.},\tilde{\gamma})}_{\mDRED}%(e_e)
%   \,.
%  \label{eq:split}
%\end{align}
As before, QCD corrections to the subprocess $\gamma^{*}\!\to\!q\bar{q}$
are closely related to the quark form factor, this time evaluated in \DRED.
Taking (4.2b) of \cite{Gnendiger:2014nxa}, identifying the UV-renormalized
couplings, and setting $\Neps\!=\!2\epsilon$ we obtain%
\footnote{In \cite{Gnendiger:2014nxa} the value of the quark form
  factor is actually given in \FDH\ which, however, happens to
  coincide with the \DRED\ result.  For later purposes, the one-loop
  result is expanded up to $\mathcal{O}(\epsilon)$.}
\begin{subequations}
\label{eq:doubleVirtualDRED}
\begin{align}
 \sigma^{(v,\gamma)}_{\mDRED}
 = \sigma^{(0,\gamma)}_{\mDRED} & \Big(\frac{\alphas}{4\pi}\Big)^{\phantom{2}}\CF\,
 \Big\{
   \!-\!\frac{4}{\epsilon^2}
   \!-\!\frac{6}{\epsilon}
   \!-\!14\!+\!\frac{7\pi^2}{3}
   \!+\!\epsilon\Big[%
    \!-\!30
    +\frac{7\pi^2}{2}
    \!+\!\frac{28\zeta(3)}{3}
    \Big]
   \!+\!\mathcal{O}(\epsilon^2)
 \Big\}\,,
 \label{eq:doubleVirtualDREDa}
 \\
\sigma^{(vv,\gamma)}_{\mDRED}
= \sigma^{(0,\gamma)}_{\mDRED} & \Big(\frac{\alphas}{4\pi}\Big)^2\CF\,\times
\notag\\*
\Big\{
C_F & \Big[
  \frac{8}{\epsilon ^4}
  \!+\!\frac{24}{\epsilon ^3}
  \!+\!\frac{
    74
    \!-\!\frac{28 \pi ^2}{3}
    }{\epsilon ^2}
  \!+\!\frac{
    \frac{401}{2}
    \!-\!26 \pi ^2
    \!-\!\frac{184 \zeta (3)}{3}
    }{\epsilon }
  \!+\!\frac{2079}{4}
  \!-\!\frac{232 \pi ^2}{3}
  \!-\!172 \zeta (3)
  \!+\!\frac{137 \pi ^4}{45}
  \Big]
\notag\\
+\,
C_A & \Big[
  \frac{11}{\epsilon ^3}
  +\frac{
    \frac{23}{9}
    \!+\!\frac{\pi ^2}{3}
    }{\epsilon ^2}
  -\frac{
    \frac{1075}{54}
    \!+\!\frac{11 \pi ^2}{6}
    \!-\!26 \zeta (3)
    }{\epsilon }
  -\frac{45943}{324}
  \!+\!\frac{535 \pi ^2}{27}
  \!+\!\frac{626 \zeta (3)}{9}
  \!-\!\frac{8 \pi ^4}{45}
  \Big]
\notag\\*
+\,N_F & \Big[
  \!-\!\frac{2}{\epsilon ^3}
  -\frac{8}{9\,\epsilon ^2}
  +\frac{
    \frac{92}{27}
    \!+\!\frac{\pi ^2}{3}
    }{\epsilon }
  +\frac{1921}{81}
  \!-\!\frac{91 \pi ^2}{27}
  \!+\!\frac{4 \zeta (3)}{9}
  \Big]
+\mathcal{O}(\epsilon)
\Big\}\,.
\label{eq:doubleVirtualDREDb}
\end{align}
\end{subequations}
The tree-level cross section is given in \eqref{eq:sigma0DRED1}.

To obtain the virtual cross section including the $\epsilon$-scalar
photon we have to determine the UV-renormalized one- and two-loop QCD
corrections to the subprocess
$\tilde{\gamma}^{*}\!\to\!q\bar{q}$. Extending the one-loop result (2.19d) of
\cite{Gnendiger:2017pys} to include the $\mathcal{O}(\epsilon)$ terms, we find
\begin{align}
 \sigma^{(v,\tilde{\gamma})}_{\mDRED}
 =\sigma^{(0,\tilde{\gamma})}_{\mDRED}
 \Big(\frac{\alphas}{4\pi}\Big)^{\phantom{2}}\CF\,
 \Big\{
   \!-\!\frac{4}{\epsilon^2}
   \!-\!\frac{6}{\epsilon}
   \!-\!10\!+\!\frac{7\pi^2}{3}
   \!+\!\epsilon\Big[%
    \!-\!16
    +\frac{7\pi^2}{3}
    \!+\!\frac{28\zeta(3)}{3}
    \Big]
   \!+\!\mathcal{O}(\epsilon^2)
 \Big\}\,,
 \label{eq:doubleVirtualDREDc}
\end{align}
with the tree-level cross section given in \eqref{eq:sigma0DRED2}.
The corresponding two-loop correction is so far unknown and has to be
determined by means of an independent computation. More precisely,
writing the cross section in terms of form-factor coefficients, we have
to compute
\begin{align}
 \sigma^{(vv,\tilde{\gamma})}_{\mDRED}
 =\sigma^{(0,\tilde{\gamma})}_{\mDRED}
  \Big[
    2 \,\tilde{F}^{(2)}_{q,\text{\DRED}}
    + \big(\tilde{F}^{(1)}_{q,\text{\DRED}}\big)^2
  \Big]\,,
  \label{eq:doubleVirtualEpsInit}
\end{align}
where $\tilde{F}^{(i)}_{q}$ is the $i$-loop form factor of the
subprocess $\tilde{\gamma}^{*}\!\to\!q\bar{q}$. It is important to
realize that in order to get the finite terms of the cross section
\eqref{eq:doubleVirtualEpsInit}, it is sufficient to compute the
\textit{divergent} part of the expression in the brackets. This is due
to the fact that $\sigma^{(0,\tilde{\gamma})}_{\mDRED}$ itself is
proportional to $\Neps\!=\!2\epsilon$.  For an arbitrary
UV-renormalized amplitude, the (process-dependent) structure of the IR
divergences is given by a $\mathbf{Z}$ matrix in colour space which is
typically given in the form
$\text{ln}\,\mathbf{Z}=(\text{ln}\,\mathbf{Z})^{(1)}+
(\text{ln}\,\mathbf{Z})^{(2)}+\mathcal{O}(\alpha^3)$.  It can be
expressed in terms of (scheme-dependent) IR anomalous dimensions
\cite{Catani:1998bh,Becher:2009cu,Gardi:2009qi,Becher:2009qa,Gardi:2009zv}
which are known in \DRED\ up to
NNLO~\mbox{\cite{Gnendiger:2014nxa,Broggio:2015ata,Broggio:2015dga}}.
Using the one-loop result of the cross section, we can then write
\begin{align}
 \sigma^{(vv,\tilde{\gamma})}_{\mDRED}
 =\ & \sigma^{(0,\tilde{\gamma})}_{\mDRED}\,
 \Big[
  2\,(\text{ln}\,\mathbf{Z})^{(2)}
  +\frac{1}{2}\,\Big(
    \sigma^{(v,\tilde{\gamma})}_{\mDRED}/
    \sigma^{(0,\tilde{\gamma})}_{\mDRED}\Big)^2
  +\mathcal{O}(\epsilon^0)
 \Big]\,.
 \label{eq:doubleVirtualEps}
\end{align}
Accordingly, it is possible to obtain the double-virtual cross section
related to the $\epsilon$-scalar photon \textit{without} performing a genuine
two-loop computation. This is not only true for the subprocess at hand but for
all processes with $\epsilon$-scalars in the initial and/or final state.

The IR structure of $\tilde{\gamma}^{*}\!\to\!q\bar{q}$ is solely governed by
the IR anomalous dimension of the (anti)quark as well as the cusp anomalous
dimension. Using (3.13), (3.14), and (6.2) of~\cite{Gnendiger:2014nxa},
identifying the renormalized couplings, and setting $\Neps\!=\!2\epsilon$,
we obtain the two-loop $\mathbf{Z}$ factor
 \begin{align}
 (\text{ln}\,\mathbf{Z})^{(2)}
 =\ & \Big(\frac{\alphas}{4\pi}\Big)^2\CF\,
\Big\{
 C_F \Big[
\frac{
    -\frac{7}{4}
    \!+\!\pi ^2
    \!-\!12\,\zeta (3)
    }{\epsilon}
  \Big]
%\notag\\*
+
%& \,
C_A \Big[
  \frac{11}{2\,\epsilon ^3}
  +\frac{
    \frac{23}{18}
    \!+\!\frac{\pi ^2}{6}
    }{\epsilon ^2}
  -\frac{
    \frac{1075}{108}
    \!+\!\frac{11 \pi ^2}{12}
    \!-\!13 \zeta (3)
    }{\epsilon}
  \Big]
\notag\\
 & \qquad\qquad\quad
+ N_F  \Big[
  \!-\!\frac{1}{\epsilon ^3}
  -\frac{4}{9\,\epsilon ^2 }
  +\frac{
    \frac{46}{27}
    \!+\!\frac{\pi ^2}{6}
    }{\epsilon}
  \Big]
  \Big\}\,.
\end{align}
Using this in \eqref{eq:doubleVirtualEps} together with
\eqref{eq:doubleVirtualDREDc}, we finally get for the cross section
\begin{align}
\sigma^{(vv,\tilde{\gamma})}_{\mDRED}
%  =\ & \sigma^{(0,\tilde{\gamma})}_{\mDRED}\,
%  \Big[
%   2\,\text{ln}\,\mathbf{Z}_{q\bar{q}}^{(2)}
%   +\frac{1}{2}\,\Big(
%     \sigma^{(v,\tilde{\gamma})}_{\mDRED}/
%     \sigma^{(0,\tilde{\gamma})}_{\mDRED}\Big)^2
%   +\mathcal{O}(\epsilon)
%  \Big]
% \\
=\ & \sigma^{(0,\tilde{\gamma})}_{\mDRED} \Big(\frac{\alphas}{4\pi}\Big)^2\CF\,
\Big\{
C_F \Big[
  \frac{8}{\epsilon ^4}
  \!+\!\frac{24}{\epsilon ^3}
  \!+\!\frac{
    58
    \!-\!\frac{28 \pi ^2}{3}
    }{\epsilon ^2 }
  \!+\!\frac{
    \frac{241}{2}
    \!-\!\frac{64 \pi ^2}{3}
    \!-\!\frac{184 \zeta (3)}{3}
    }{\epsilon}
  \Big]
\notag\\*
+ & \,
C_A \Big[
  \frac{11}{\epsilon ^3}
  \!+\!\frac{
    \frac{23}{9}
    \!+\!\frac{\pi ^2}{3}
    }{\epsilon ^2}
  \!-\!\frac{
    \frac{1075}{54}
    \!+\!\frac{11 \pi ^2}{6}
    \!-\!26 \zeta (3)
    }{\epsilon}
  \Big]
%\notag\\
+ N_F  \Big[
  \!-\!\frac{2}{\epsilon ^3}
  \!-\!\frac{8}{9\,\epsilon ^2 }
  \!+\!\frac{
    \frac{92}{27}
    \!+\!\frac{\pi ^2}{3}
    }{\epsilon}
  \Big]
%\notag\\*
+  \mathcal{O}(\epsilon^0)
\Big\}\,.
\phantom{\frac{1}{1}}
\label{eq:doubleVirtualDREDd}
\end{align}
% \end{subequations}

As a cross check of our result, we note that the $\mathbf{Z}$ factors (but
\textit{not} the form factors) related to $\tilde{\gamma}^{*}\!\to\!q\bar{q}$
and $\gamma^{*}\!\to\!q\bar{q}$ are exactly the same. Writing down
\eqref{eq:doubleVirtualEps} for each process separately, it is possible to
eliminate $\text{ln}\,\mathbf{Z}$, i.\,e.\ to write down the identity
\begin{align}
 \Bigg[
 \frac{\sigma^{(vv,\tilde{\gamma})}_{\mDRED}}
  {\sigma^{(0,\tilde{\gamma})}_{\mDRED}}
 -\frac{1}{2}\bigg(
  \frac{\sigma^{(v,\tilde{\gamma})}_{\mDRED}}
  {\sigma^{(0,\tilde{\gamma})}_{\mDRED}}
  \bigg)^2
  \Bigg]_{\text{div.}}
 =
 \Bigg[
 \frac{\sigma^{(vv,\gamma)}_{\mDRED}}
  {\sigma^{(0,\gamma)}_{\mDRED}}
 -\frac{1}{2}\bigg(
  \frac{\sigma^{(v,\gamma)}_{\mDRED}}
  {\sigma^{(0,\gamma)}_{\mDRED}}
  \bigg)^2
  \Bigg]_{\text{div.}}
  \label{eq:IRrel}
\end{align}
for the IR divergences. We have checked explicitly that this relation holds.

Of course, it is also possible to obtain $\sigma^{(vv,\tilde{\gamma})}_{\mDRED}$
in the usual way by explicitly computing the virtual two-loop corrections to
the subprocess $\tilde{\gamma}^{*}\!\to\!q\bar{q}$. Compared to \CDR, one has to
consider the UV renormalization of the evanescent couplings in the following way:
 \begin{itemize}
  \item QED:
     In contrast to the electromagnetic gauge coupling~$e$
     which does not get renormalized due to a Ward identity, the UV
     renormalization of the evanescent coupling~$e_e$ has to be
     included. As the (bare) coupling is already present at the tree-level,
     its renormalization has to be known up to two loops.%
  \item QCD:
     The different renormalization of the strong gauge coupling~$g_s$
     and the corresponding evanescent coupling~$g_e$ has to be considered.%
     \footnote{The coupling~$g_e$ mediates the interaction of fermions
       and $\epsilon$-scalar gluons. The latter originate from splitting quasi
       $\dim_s$-dimensional gluon fields similar to~\eqref{eq:vectorFieldSplit}.}
   As the (bare) QCD couplings contribute starting from one loop,
   their renormalization has to be known up to the one-loop level as well.
 \end{itemize}
 The renormalization of the QCD couplings $g_s$ and $g_e$ is well
 known~\cite{Harlander:2006rj}. It already had to be considered in the
 determination of \eqref{eq:doubleVirtualDREDb} to obtain the correct result,
 see~\cite{Gnendiger:2014nxa} for more details.
 The renormalization of $e_e$ is so far only known at one-loop~%
 \cite{Gnendiger:2017pys}. At two loops it has to be determined in a
 separate step, either by using  generalized renormalization group
 equations~\cite{Luo:2002ti} or by  direct computation. Identifying the
 renormalized QCD couplings and setting $\Neps\!=\!2\epsilon$, we find
 for the $\MS$ renormalization $Q_q\, e_e \to Q_q \, Z_{e_e} e_e $
  \begin{align}
    Z_{e_e}=
     %\Big[
       1
       & +\Big(\frac{\alphas}{4\pi}\Big)\,
 	\CF\,\Big[
 	  \!-\!\frac{1}{\epsilon}
 	  \!-\!1
 	  +\mathcal{O}(\epsilon)
 	  \Big]
 \notag\\*
       &+\Big(\frac{\alphas}{4\pi}\Big)^2\,C_F\,\Big\{
 	C_F\,\Big[
 	  \!-\!\frac{3}{2\,\epsilon^2}
 	  \!+\!\frac{13}{4\,\epsilon}
 	  \Big]
 	\!+\!C_A\,\Big[
 	  \frac{7}{2\,\epsilon^2}
 	  \!-\!\frac{19}{12\,\epsilon}
 	  \Big]
 	\!-\!N_F\,\Big[
 	  \frac{2}{3\,\epsilon}
 	  \Big]
 	+\mathcal{O}(\epsilon^0)
     \Big\}
    %\Big]
 %\notag\\*&
    +\mathcal{O}(\alphas^3)
    \,.
    \phantom{\frac{1}{1}}
  \end{align}
We would like to emphasize that the two-loop renormalization of $e_e$ only
has to be known if the double-virtual contribution is obtained via an explicit
two-loop computation. In contrast, when using the \eqref{eq:doubleVirtualEps}
for the determination of the cross section it only has to be known at one-loop.

Finally, we stress that in order to obtain the physical NNLO result of the
cross section $e^{+}e^{-}\!\to\gamma^{*}\!\to\!2~jets$ it is in principle
sufficient to only use the virtual corrections in \eqref{eq:doubleVirtualDRED}
since all contributions related to the evanescent coupling $e_e$ drop
out in the final result. This will be shown explicitly in
Section\,\ref{sec:epsContributions}.  In the next section, however,
we describe an efficient method for the determination of the real
corrections where the evanescent degrees of freedom are not
distinguished but automatically included.  To obtain the physical
cross section with this approach one therefore needs the virtual
contributions in \eqref{eq:doubleVirtualDREDc} and
\eqref{eq:doubleVirtualDREDd}.\\

\subsection{Real corrections}
\label{sec:doubleReal}

To obtain the double-real corrections we have to integrate the squared
tree-level matrix elements of the processes
$e^{+}e^{-}\!\to\gamma^{*}\!\to\!q\bar{q}\,g g$,
$e^{+}e^{-}\!\to\gamma^{*}\!\to\!q\bar{q}\,q\bar{q}$, and
$e^{+}e^{-}\!\to\gamma^{*}\!\to\!q\bar{q}\,q'\bar{q}'$ over the phase
space.  The standard procedure in \CDR\ is to compute the matrix
elements in $\dim$ dimensions, leading to terms of
$\mathcal{O}(\epsilon^m)$ with $m\!>\!0$.  Due to IR singularities of
the form $1/\epsilon^n$ with $n\!\le\!4$ in the phase-space
integration, these terms can not be neglected.  For the case at hand,
it is possible to do the complete phase-space integration
analytically, as discussed in~\cite{Gehrmann-DeRidder:2003pne}. These
results have been used in~\cite{GehrmannDeRidder:2004tv} to compute
the double-real corrections in \CDR. Pulling out the same prefactor as
for the virtual contributions, they read
\begin{subequations}
\begin{align}
\sigma^{(r)}_{\mCDR}
=  \sigma&^{(0)}_{\mCDR}
  \Big(\frac{\alphas}{4\pi}\Big)^{2}\CF\,
 \Big\{
   \frac{4}{\epsilon^2}
   +\frac{6}{\epsilon}
   +19
   -\frac{7\pi^2}{3}
   +\mathcal{O}(\epsilon)
 \Big\}\,,
\\
\sigma^{(rr)}_{\mCDR}
=  \sigma&^{(0)}_{\mCDR}
  \Big(\frac{\alphas}{4\pi}\Big)^{2}\CF\,\times
  %\Phi_2(\epsilon)\,s^{-\epsilon}\,
  %c_\Gamma(\epsilon)\,
\notag\\*
\Big\{
C_F & \Big[
  \frac{8}{\epsilon ^4}
  +\frac{24}{\epsilon ^3}
  +\frac{
    104
    \!-\!12 \pi ^2
    }{\epsilon ^2}
  +\frac{
    \frac{819}{2}
    \!-\!34 \pi ^2
    \!-\!\frac{664 \zeta (3)}{3}
    }{\epsilon }
  +\frac{6243}{4}
  \!-\!\frac{439 \pi ^2}{3}
  \!-\!556 \zeta (3)
  \!+\!\frac{53 \pi ^4}{45}
  \Big]
\notag\\*
+\,
 C_A & \Big[
  \frac{2}{\epsilon ^4}
  \!+\!\frac{29}{3\,\epsilon ^3}
  \!+\!\frac{
    \frac{400}{9}
    \!-\!\frac{8 \pi ^2}{3}
    }{\epsilon ^2}
  \!+\!\frac{
    \frac{10555}{54}
    \!-\!\frac{283 \pi ^2}{18}
    \!-\!\frac{94 \zeta (3)}{3}
    }{\epsilon }
  \!+\!\frac{285517}{324}
  \!-\!\frac{2002 \pi ^2}{27}
  \!-\!\frac{2806 \zeta (3)}{9}
  \!+\!\frac{89 \pi ^4}{60}
  \Big]
\notag\\*
+\,N_F & \Big[
  \!-\!\frac{2}{3\,\epsilon ^3}
  -\frac{28}{9\,\epsilon ^2}
  -\frac{
    \frac{407}{27}
    \!-\!\frac{11 \pi ^2}{9}
    }{\epsilon }
  -\frac{11753}{162}
  \!+\!\frac{154 \pi ^2}{27}
  \!+\!\frac{268 \zeta (3)}{9}
  \Big]
+\mathcal{O}(\epsilon)
\Big\}\,.
\end{align}
\end{subequations}

One main statement of this work is that the double-real corrections in
\DRED\ are obtained simply by computing the squared real matrix element in
four dimensions, i.\,e\ by setting $\dim_s\!=\!4$ \textit{throughout}
the computation.  The subsequent phase-space integration is done in the
usual way, i.\,e.\ in $\dim$ dimensions. As for loop integrations, this
regularization is mandatory since setting $d\!\to\!4$ would lead to ill-defined
expressions and, consequently, the phase-space integrals are precisely those
computed in~\cite{Gehrmann-DeRidder:2003pne}.
Using the code of~\cite{GehrmannDeRidder:2004tv}, we obtain
%\subsection*{DRED}
\begin{subequations}
\begin{align}
 \sigma^{(r)}_{\mDRED}
=  \sigma&^{(0)}_{\mDRED}
\Big(\frac{\alphas}{4\pi}\Big)^{2}\CF\,
 \Big\{
   \frac{4}{\epsilon^2}
   +\frac{6}{\epsilon}
   +17
   -\frac{7\pi^2}{3}
   +\mathcal{O}(\epsilon)
 \Big\}\,,
 \label{eq:doubleRealDREDa}
\\
 \sigma^{(rr)}_{\mDRED}
 = \sigma&^{(0)}_{\mDRED}
   \Big(\frac{\alphas}{4\pi}\Big)^2\CF\,
   %c_\Gamma^2(\epsilon)\,
   %\Phi_2(\epsilon)\,s^{-\epsilon}
   \times\,
 \notag\\*
 \Big\{
  C_F & \Big[
   \frac{8}{\epsilon^4}
   \!+\!\frac{24}{\epsilon^3}
   \!+\!\frac{
     96
     \!-\!12\pi^2
     }{\epsilon^2}
   +\frac{
     \frac{743}{2}
     \!-\!34\pi^2
     \!-\!\frac{664\zeta(3)}{3}
     }{\epsilon}
   +\frac{5587}{4}
   \!-\!135\pi^2
   \!-\!556\zeta(3)
   \!+\!\frac{53\pi^4}{45}
   \Big]
 \notag\\
 + \,
  C_A & \Big[
   \frac{2}{\epsilon^4}
   \!+\!\frac{29}{3\,\epsilon^3}
   \!+\!\frac{
     \frac{379}{9}
     \!-\!\frac{8\pi^2}{3}
     }{\epsilon^2}
   \!+\!\frac{
     \frac{10021}{54}
     \!-\!\frac{283\pi^2}{18}
     \!-\!\frac{94\zeta(3)}{3}
   }{\epsilon}
   \!+\!\frac{272851}{324}
   \!-\!\frac{3809\pi^2}{54}
   \!-\!\frac{2806\zeta(3)}{9}
   \!+\!\frac{89\pi^4}{60}
   \Big]
 \notag\\
 + \, N_F & \Big[
   \!-\!\frac{2}{3\,\epsilon^3}
   -\frac{28}{9\,\epsilon^2}
   -\frac{
     \frac{380}{27}
     \!-\!\frac{11\pi^2}{9}
     }{\epsilon}
   -\frac{5296}{81}
   \!+\!\frac{154\pi^2}{27}
   \!+\!\frac{268\zeta(3)}{9}
   \Big]
 +\mathcal{O}(\epsilon)
 \Big\}\, ,
\end{align}
\end{subequations}
with the tree-level cross section given in \eqref{eq:bornRel}.
The two leading poles in the curly brackets agree with the \CDR\ result
whereas the subsequent terms differ. As we will see, these differences
cancel for the physical cross section.

In the particular case at hand, using a four-dimensional algebra for
the double-real corrections implies setting $d\!\to\!4$ in the
\CDR\ matrix element, combined with the usual phase-space integration in
$\dim$ dimensions.  More complicated processes are usually treated by
subtracting the singular limits of the matrix elements and by adding
them back as IR 'counterterms'. The subtracted matrix element is
always treated in four dimensions, as by construction it is finite
upon integration. The statement regarding \DRED\ in this case is that
also the integrand of the IR counterterm is required in four
dimensions only.\\

\subsection{Real-virtual corrections}
\label{sec:realVirtual}

To obtain the real-virtual corrections, the algebra for the process
$e^{+}e^{-}\!\to\gamma^{*}\!\to\!q\bar{q}\,g$ has to be carried out.
This amounts to calculating $2\, \text{Re} \langle\cA^{(0)}|\cA^{(1)}\rangle$
where $\cA^{(0)}$ and $\cA^{(1)}$ are the tree-level and the bare one-loop
amplitudes for this process, respectively.
In \CDR, the Lorentz algebra is obviously carried out in $\dim$ dimensions.
Writing dot products containing the loop momentum in terms of denominators,
the result can be expressed as a sum of coefficients times scalar loop integrals.
In practice, we use integration-by-parts identities as implemented in FIRE~%
\cite{Smirnov:2008iw} to reduce all loop integrals to master integrals
which in turn are expressed in terms of dot products of external momenta.
The intermediate result contains $1/\epsilon^n$ poles with $n\!\le\!2$ and is
finally integrated over the massless three-particle phase space.

To obtain the renormalized real-virtual contribution we have to add counterterms.
In \CDR, the only counterterm that contributes is the one induced through the
coupling renormalization of $M^{(0)}$.
Considering this and putting everything together, we obtain
\begin{align}
\sigma^{(rv)}_{\mCDR}
=\ \sigma&^{(0)}_{\mCDR} \Big(\frac{\alphas}{4\pi}\Big)^{2}\CF\,
  %\Phi_2(\epsilon)\,s^{-\epsilon}\,
  \times
  %c_\Gamma(\epsilon)\,
\notag\\*
\Big\{
 C_F & \Big[
  \!-\!\frac{16}{\epsilon ^4}
  -\frac{48}{\epsilon ^3}
  -\frac{
    186
    \!-\!\frac{64 \pi ^2}{3}
    }{\epsilon ^2}
  -\frac{
    632
    \!-\!60 \pi ^2
    \!-\!\frac{848 \zeta (3)}{3}
    }{\epsilon }
  -2138
  +\frac{697 \pi ^2}{3}
  \!+\!728 \zeta (3)
  \!-\!\frac{38 \pi ^4}{9}
  \Big]
\notag\\
+
C_A & \Big[
 \!-\!\frac{2}{\epsilon ^4}
 -\frac{62}{3\,\epsilon ^3}
 -\frac{
  48
  \!-\!\frac{7 \pi ^2}{3}
  }{\epsilon ^2}
 -\frac{
  \frac{533}{3}
  \!-\!\frac{158 \pi ^2}{9}
  \!-\!\frac{16 \zeta (3)}{3}
  }{\epsilon }
  -\frac{3971}{6}
  \!+\!\frac{109 \pi ^2}{2}
  \!+\!\frac{1784 \zeta (3)}{9}
  \!-\!\frac{47 \pi ^4}{36}
  \Big]
\notag\\
+\,N_F  & \Big[
  \frac{8}{3\,\epsilon ^3}
  +\frac{4}{\epsilon ^2}
  +\frac{
    \frac{38}{3}
    \!-\!\frac{14 \pi ^2}{9}
    }{\epsilon }
  +\frac{109}{3}
  \!-\!\frac{7 \pi ^2}{3}
  \!-\!\frac{200 \zeta (3)}{9}
  \Big]
  +\mathcal{O}(\epsilon)
\Big\} \, ,
\end{align}
in agreement with~\cite{GehrmannDeRidder:2004tv}.
 
The first difference in the \DRED\ computation is that the algebra is done
in $\dim_s\!=\!4$ dimensions, similar to the double-real contributions.
More precisely, in \DRED\ the coefficients that multiply the scalar loop
integrals are obtained from the corresponding coefficients in \CDR\ by
setting $\dim\!\to\!4$. All subsequent steps of the computation
(integration-by-parts, phase-space integration) are again performed in the
usual way, i.\,e.\ in $d$ dimensions.  A second difference between
\DRED\ and \CDR\ concerns renormalization. While for the evaluation of
the bare result it is not necessary to distinguish the various couplings
and subprocesses, the proper UV renormalization requires a careful
distinction between them, similar to the virtual contributions.
More concretely, the real-virtual contributions in \DRED\ have to be
renormalized by adding the counterterm
\begin{align}
\label{ren:rvdred}
\delta \sigma^{(rv)}_{\mDRED} &
=2\Big\{
  \delta Z_{g_s}^{(1)}\,
  \sigma^{(0)}_{\gamma\to q\bar{q} g}
\!+\! \delta Z_{g_e}^{(1)}\,
  \sigma^{(0)}_{\gamma\to q\bar{q} \tilde{g}}
\!+\! \big[\delta Z_{e_e}^{(1)}\! +\! \delta Z_{g_s}^{(1)}\big]\,
  \sigma^{(0)}_{\tilde{\gamma}\to q\bar{q}g}
\!+\! \big[\delta Z_{e_e}^{(1)}\! +\! \delta Z_{g_e}^{(1)} \big]\,
  \sigma^{(0)}_{\tilde{\gamma}\to q\bar{q}\tilde{g}}
  \Big\}\, ,
%\phantom{\Bigg|}
\end{align}
where $\tilde{g}$ and $\delta Z_i^{(1)}$ denote an $\epsilon$-scalar gluon and
the NLO part of the renormalization factor $Z_i$ for the various couplings,
respectively. The first term in \eqref{ren:rvdred} is precisely the counterterm
in \CDR, where no renormalization of the electromagnetic coupling $e$ is required.
The other terms are present due to the subprocesses including $\epsilon$-scalar
photons and/or $\epsilon$-scalar gluons, see Section~\ref{sec:doubleVirtual} for
more details.

Adding \eqref{ren:rvdred} to the bare result, dropping the distinction
between the renormalized gauge and scalar couplings, and setting
$\Neps\!=\!2\epsilon$ we obtain
\begin{align}
\sigma^{(rv)}_{\mDRED}
= \sigma&^{(0)}_{\mDRED} \Big(\frac{\alphas}{4\pi}\Big)^{2}\CF\,
  %\Phi_2(\epsilon)\,s^{-\epsilon}\,
  \times
  %c_\Gamma(\epsilon)\,
\notag\\*
\Big\{
C_F & \Big[
  \!-\!\frac{16}{\epsilon ^4}
  \!-\!\frac{48}{\epsilon ^3}
  \!-\!\frac{
    170
    \!-\!\frac{64 \pi ^2}{3}
    }{\epsilon ^2}
  \!-\!\frac{
    556
    \!-\!60 \pi ^2
    \!-\!\frac{848 \zeta (3)}{3}
    }{\epsilon }
  \!-\!1838
  \!+\!\frac{623 \pi ^2}{3}
  \!+\!728 \zeta (3)
  \!-\!\frac{38 \pi ^4}{9}
  \Big]
\notag\\
+\,
 C_A & \Big[
  \!-\!\frac{2}{\epsilon ^4}
  \!-\!\frac{62}{3\,\epsilon ^3}
  \!-\!\frac{
    \frac{134}{3}
    \!-\!\frac{7 \pi ^2}{3}
    }{\epsilon ^2}
  \!-\!\frac{
    \frac{497}{3}
    \!-\!\frac{158 \pi ^2}{9}
    \!-\!\frac{16 \zeta (3)}{3}
    }{\epsilon }
  \!-\!\frac{3833}{6}
  \!+\!\frac{913 \pi ^2}{18}
  \!+\!\frac{1784 \zeta (3)}{9}
  \!-\!\frac{47 \pi ^4}{36}
  \Big]
\notag\\* 
+ \, N_F & \Big[
  \frac{8}{3\, \epsilon ^3}
  +\frac{4}{\epsilon ^2}
  +\frac{
    \frac{32}{3}
    \!-\!\frac{14 \pi ^2}{9}
    }{\epsilon }
  +\frac{92}{3}
  \!-\!\frac{7 \pi ^2}{3}
  \!-\!\frac{200 \zeta (3)}{9}
  \Big]
  +\mathcal{O}(\epsilon)
\Big\}\,.
\end{align}
Similar to the double-virtual and double-real corrections, the two
leading poles in the curly brackets agree with the \CDR\ result.\\

\subsection{Combination of the contributions}
\label{sec:combination}

Adding the double-virtual, the double-real, and the real-virtual
contributions we finally get the well-known result for the NNLO
prediction~\cite{Celmaster:1979xr, Chetyrkin:1979bj}, namely
\begin{subequations}
\label{eq:finalRes}
\begin{align}
 \sigma^{(2)}
&=\Big[
\sigma^{(0)}_{\text{\CDR}}
+\sigma^{(v)}_{\text{\CDR}}
+\sigma^{(r)}_{\text{\CDR}}
+\sigma^{(vv)}_{\text{\CDR}}
+\sigma^{(rr)}_{\text{\CDR}}
+\sigma^{(rv)}_{\text{\CDR}}
\Big]_{d\to 4}
\phantom{\bigg|}
\\*
&\equiv\Big[
\sigma^{(0)}_{\text{\DRED}}
+\sigma^{(v)}_{\text{\DRED}}
+\sigma^{(r)}_{\text{\DRED}}
+\sigma^{(vv)}_{\text{\DRED}}
+\sigma^{(rr)}_{\text{\DRED}}
+\sigma^{(rv)}_{\text{\DRED}}
\Big]_{d\to 4}
\phantom{\bigg|}
\label{eq:dredall}
\\*[5pt]
&=  %\frac{\eqSq\,N_c}{3\, s}\Big(\frac{e^4}{4\pi}\Big)
\sigma^{(0)}
\Big[\,
  1
  \!+\!\Big(\frac{\alpha_s}{4\pi}\Big) \, 3\, \CF
  \!+\!\Big(\frac{\alpha_s}{4\pi}\Big)^2 \CF \Big\{
    C_F\Big[
    \!-\!\frac{3}{2}
    \Big]
    \!+\!C_A\Big[
     \frac{123}{2}
     \!-\!44\zeta(3)
     \Big]
    \!+\!N_F\Big[
      \!-\!11
      \!+\!8\zeta(3)
      \Big]
  \Big\}
\Big].
\label{fdh:xs}
\end{align}
\end{subequations}
Also at NNLO, the physical cross section is therefore the same in
\CDR\ and \DRED, confirming that \DRED\ can be used consistently for
real corrections at NNLO. Using the four-dimensional approach for the
determination of the purely real contributions as described in
Section~\ref{sec:doubleReal}, we have to identify the renormalized
couplings of QED ($e_e$ and~$e$) and QCD ($g_e$ and $g_s$),
respectively, and have to set $\Neps\!=\!2\epsilon$ at the end of the
calculation.\\

\subsection{Contributions from $\epsilon$-scalar photons}
\label{sec:epsContributions}

As mentioned repeatedly, the evanescent couplings in \DRED\ are in
principle independent of the corresponding gauge couplings and for
technical reasons it is often advantageous to set equal their
renormalized values. This is possible since physical cross sections
are actually independent of the evanescent couplings. Hence, their
value is irrelevant. This can be understood by noting that
contributions involving evanescent couplings are due to processes with
$\epsilon$-scalar photons~$\tilde{\gamma}$ and/or $\epsilon$-scalar
gluons~$\tilde{g}$.  As for normal gauge bosons, cross sections with
$\epsilon$-scalars are finite, i.\,e.\ they contain no $1/\epsilon^n$
poles.  They come, however, with a multiplicity of $\Neps$, are
therefore of
$\mathcal{O}(\Neps/\epsilon^0)\!=\!\mathcal{O}(\epsilon)$, and vanish
in the physical limit $\dim\!\to\!4$. In the remainder of this section
we explicitly show this for the process with an $\epsilon$-scalar
photon $\tilde{\gamma}$, i.\,e.\ for the terms proportional to the
coupling~$e_e$.

Starting at NLO, the virtual contributions for the exchange of
$\tilde{\gamma}$ are given in \eqref{eq:doubleVirtualDREDc};
the corresponding real contributions are included in
\eqref{eq:doubleRealDREDa}.
To obtain $\sigma^{(r,\tilde{\gamma})}_{\mDRED}$ separately,
the squared tree-level matrix element for the process
$e^{+}e^{-}\!\to\tilde{\gamma}^{*}\!\to\!q\bar{q}\,g$ has to be
integrated over the three-particle phase space. We find
\begin{align}
   \sigma^{(r,\tilde{\gamma})}_{\mDRED}
 = \sigma^{(0,\tilde{\gamma})}_{\mDRED} &
 \Big(\frac{\alphas}{4\pi}\Big)^{\phantom{2}}\CF\,
 \Big\{
   \frac{4}{\epsilon^2}
   +\frac{6}{\epsilon}
   +\mathcal{O}(\epsilon^0)
 \Big\}\,,
 \label{eq:RealDRED2}
\end{align}
with the tree-level cross section given in \eqref{eq:sigma0DRED2}.
Not surprisingly, the poles in $ \sigma^{(v,\tilde{\gamma})}_{\mDRED}
+ \sigma^{(r,\tilde{\gamma})}_{\mDRED}$ cancel and we are left with a
contribution $\propto\Neps/\epsilon^0$ that vanishes in the limit
$d\!\to\!4$.

Moving to NNLO, we again need the contributions from the $\epsilon$-scalar
photon separately. For the virtual part they are given in
\eqref{eq:doubleVirtualDREDd}; for the double-real and the
real-virtual part we write 
 \begin{align}
   \sigma^{(rr)}_{\mDRED}
    &=\sigma^{(rr,\gamma)}_{\mDRED}
    +\sigma^{(rr,\tilde{\gamma})}_{\mDRED}\,,
   %\\
   \qquad
   \sigma^{(rv)}_{\mDRED}
    =\sigma^{(rv,\gamma)}_{\mDRED}
    +\sigma^{(rv,\tilde{\gamma})}_{\mDRED}
 \end{align}
and note that in order to obtain
$\sigma^{(rr,\tilde{\gamma})}_{\mDRED}$ and
$\sigma^{(rv,\tilde{\gamma})}_{\mDRED}$ a separate computation is
required. Carrying out this calculation as described in
Sections~\ref{sec:doubleReal} and \ref{sec:realVirtual}, we obtain
\begin{subequations}
\begin{align}
\sigma^{(rr,\tilde{\gamma})}_{\mDRED}
= &\, \sigma^{(0,\tilde{\gamma})}_{\mDRED}
\Big(\frac{\alphas}{4\pi}\Big)^{2}\CF\,
\Big\{
C_F \Big[
  \frac{8}{\epsilon ^4}
  \!+\!\frac{24}{\epsilon ^3}
  \!+\!\frac{
    96
    \!-\!12 \pi ^2
    }{\epsilon ^2}
  \!+\!\frac{
    \frac{743}{2}
    \!-\!34 \pi ^2
    \!-\!\frac{664 \zeta (3)}{3}
    }{\epsilon}
  \Big]
\notag\\*
+ & \,
C_A \Big[
  \frac{2}{\epsilon ^4}
  \!+\!\frac{29}{3\,\epsilon ^3}
  \!+\!\frac{
    \frac{379}{9}
    \!-\!\frac{8 \pi ^2}{3}
    }{\epsilon ^2}
  \!+\!\frac{
    \frac{9985}{54}
    \!-\!\frac{283 \pi ^2}{18}
    \!-\!\frac{94 \zeta (3)}{3}
    }{\epsilon}
  \Big]
%\notag\\*
-
N_F \Big[\,
  \frac{2}{3\,\epsilon ^3}
  \!+\!\frac{28}{9\,\epsilon ^2 }
  \!+\!\frac{
    \frac{362}{27}
    \!-\!\frac{11 \pi ^2}{9}
    }{\epsilon}
  \Big]
  \Big\}\,,
\\
\sigma^{(rv,\tilde{\gamma})}_{\mDRED}
= & \, \sigma^{(0,\tilde{\gamma})}_{\mDRED}
  \Big(\frac{\alphas}{4\pi}\Big)^{2}\CF\,
  %\Phi_2(\epsilon)\,s^{-\epsilon}\,
%\times\notag\\
\Big\{
C_F \Big[
  \!-\!\frac{16}{\epsilon ^4}
  \!-\!\frac{48}{\epsilon ^3}
  \!-\!\frac{
    154
    \!-\!\frac{64 \pi ^2}{3}
    }{\epsilon ^2}
  \!-\!\frac{
    492
    \!-\!\frac{166 \pi ^2}{3}
    \!-\!\frac{848 \zeta (3)}{3}
    }{\epsilon}
  \Big]
\notag\\*
+ & \,
C_A  \Big[
  \!-\!\frac{2}{\epsilon ^4}
  \!-\!\frac{62}{3\,\epsilon ^3}
  \!-\!\frac{
    \frac{134}{3}
    \!-\!\frac{7\pi ^2}{3}
    }{\epsilon^2 }
  \!-\!\frac{
    165
    \!-\!\frac{158 \pi ^2}{9}
    \!-\!\frac{16 \zeta (3)}{3}
    }{\epsilon}
  \Big]
+
N_F \Big[
  \frac{8}{3\,\epsilon ^3}
  \!+\!\frac{4}{\epsilon ^2 }
    \!+\!\frac{
    10
    \!-\!\frac{14 \pi ^2}{9}
    }{\epsilon}
  \Big]
%\notag\\*
%\!+\!\mathcal{O}(\epsilon^0)
  \Big\}\,.
  \phantom{\bigg{|}}
  \label{eq:epsScalarRes}
\end{align}
\end{subequations}
Combining this with \eqref{eq:doubleVirtualDREDd} one finds that the
total two-loop contribution from \mbox{$\epsilon$-scalar} photons,
$\sigma^{(2,\tilde\gamma)}_{\mDRED}$, is indeed
$\propto\!\sigma^{(0,\tilde{\gamma})}_{\mDRED}\! \times
\!\mathcal{O}(\epsilon^0) =\mathcal{O}(\epsilon)$.  Hence, also at
NNLO all terms $\propto\!e_e$ drop out in the physical limit
$d\!\to\!4$.  A similar analysis splitting the strong coupling
contributions into $g_s$ and $g_e$ parts would reveal that
the latter drop out in physical cross sections as well.

Obviously, disentangling evanescent contributions is rather
cumbersome. We should stress again that this is done here only for
illustrative purposes and is not required for the actual
computation. The main point is that it is easier to compute
e.\,g.\ $\sigma^{(rr)}_{\mDRED}$ rather than
$\sigma^{(rr,\gamma)}_{\mDRED}$ or
$\sigma^{(rr,\tilde{\gamma})}_{\mDRED}$ separately. Thus, while the
actual value of the evanescent coupling is irrelevant, for technical
reasons it can be useful to set them equal to the corresponding gauge
couplings.\\

\section{NNLO corrections in FDH and HV}
\label{sec:fdh}

\FDH\ is well adapted to several techniques for computing virtual corrections
and, hence, is often used for calculations of one- and two-loop matrix elements.
Compared to \DRED, \FDH\ is actually slightly more convenient for virtual
corrections since 'regular' vector fields are not regularized and no associated
$\epsilon$-scalar has to be considered.  Also UV renormalization is somewhat
simpler even though it is still necessary to distinguish gauge and evanescent
couplings.

In order to use \FDH~results at NNLO, either the two-loop amplitudes
have to be converted to \CDR\ (or \DRED) or the real corrections have
to be evaluated in \FDH\ as well. The first option is well understood~%
\cite{Kilgore:2012tb, Broggio:2015dga, Gnendiger:2016cpg}.
In this section we follow the second option and show that using
\FDH\ throughout the calculation results in the same total cross section,
as required. This is actually closely related to the vanishing of the
contribution due to processes with an $\epsilon$-scalar photon, as discussed
in Section~\ref{sec:epsContributions}. Indeed, the difference between
\DRED\ and \FDH\ is in the treatment of the regular vector fields.
In \FDH\ they are strictly four-dimensional, whereas in \DRED\ they are
quasi $\dim_s$-dimensional and typically split into a quasi $\dim$- and
a quasi $\Neps$-dimensional part, as in Section~\ref{sec:process}. 
For the process under consideration, the photon is the only regular
boson. Thus, in order to convert the \DRED\ results to \FDH, we simply have
to pick the \DRED\ contributions of a regular $\dim$-dimensional photon
and correct by an overall factor to effectively treat the photon in strictly
four dimensions.

To illustrate this, we start with the double-virtual contributions.
They have been evaluated in \FDH\ in~\cite{Gnendiger:2014nxa} and can be
read off from \eqref{eq:doubleVirtualDREDb} as
\begin{align}
  \label{eq:fdhvv}
  \sigma^{(vv)}_{\mFDH} = \sigma^{(vv,\gamma)}_{\mDRED}\
  \frac{\sigma^{(0)}_{\mFDH}}{\sigma^{(0,\gamma)}_{\mDRED}}
  = \frac{\sigma^{(vv,\gamma)}_{\mDRED}}{1-\epsilon} \, ,
\end{align}
where $\sigma^{(0)}_{\mFDH}\! =\! \sigma^{(0)}_{\mDRED}
\!=\! \sigma^{(0,\gamma)}_{\mDRED}\!+\! \sigma^{(0,\tilde{\gamma})}_{\mDRED}$
is the \FDH\ tree-level result, see also (2.14) of \cite{Gnendiger:2017pys}.
An analogous relation holds for the real contributions and in fact
separately for all terms contributing in \eqref{eq:finalRes}.
Hence we obtain
\begin{align}
  \label{eq:dredvsfdh}
  \sigma^{(2)}_{\mFDH} \Big|_{d\to4} 
  = \frac{\sigma^{(2,\gamma)}_{\mDRED}}{1-\epsilon}\bigg|_{d\to4}
  =  \sigma^{(2,\gamma)}_{\mDRED}\Big|_{d\to4}
  =  \sigma^{(2)}_{\mDRED}\Big|_{d\to4}
  = \sigma^{(2)} \,,
\end{align}
where we have used that $\sigma^{(2,\gamma)}_{\mDRED}$ is finite and
$\sigma^{(2,\tilde\gamma)}_{\mDRED}\big|_{d\to4}\!=\! 0$.

It should be noted that the actual computation in \FDH\ is rather
cumbersome. Regarding the virtual corrections, we first mention that
proper renormalization requires a distinction of $g_s$ and
$g_e$, as in \DRED. In addition, the regular four-dimensional
photon results in terms $\propto g^{\mu\nu}_{[4]}$ which in turn give
rise to integrals with a \textit{strictly} four-dimensional loop momentum
$k_{[4]}^2$ in the numerator.%
\footnote{In contrast, performing the algebra in \DRED\ always results
in $\dim$-dimensional loop momenta. This is ensured by the fact that
a contraction of a $\dim_s$- with a $\dim$-dimensional quantity yields
a projection onto the $\dim$-dimensional subspace, see
\cite{Gnendiger:2017pys} for more details.}
This requires a careful implementation of an algebra with $d_s$-, $d$-,
and four-dimensional objects in the one-loop
amplitude~\cite{Fazio:2014xea,Gnendiger:2017rfh}.
Regarding the double-real contributions the situation is even worse.
Even if the final-state particles are treated as 'singular',
i.\,e.\ in $d_s$ dimensions, simply integrating the matrix element
squared over phase space does not yield the correct result in
\FDH. This is related to unitarity violations~\cite{Catani:1996pk,
  Signer:2008va}.

\begin{figure}[t]
  \begin{center}
    \includegraphics[width=0.5\textwidth]{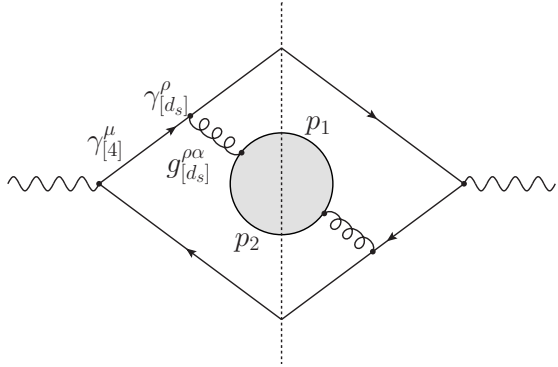}
 \end{center}
  \caption{Double-real contributions that require special treatment in
    \FDH\ to correct for unitarity violations. The blob represents
    either a quark, a gluon, or a ghost loop.}
  \label{fig:FdiagRR}
  \end{figure}

For our process such violations are restricted to the diagram shown in
Figure~\ref{fig:FdiagRR}. From a practical point of view, the problem
can be understood by noting that evaluating the trace of the inner loop
(either a quark, a gluon, or a ghost loop) gives rise to terms such as
$p_1^\alpha\,p_2^\beta$, where $p_1$ and $p_2$ are the particle momenta
and $\alpha$ and $\beta$ are the Lorentz indices of the inner loop.
In order to evaluate these diagrams correctly, the angular average
has to be performed in $d$ dimensions using identities like
\begin{align}
  p_{1}^{\alpha}\,p_{2}^{\beta} &\to
  \frac{1}{4(d\!-\!1)} \Big[
    (d\!-\!2)\, (p_{12})^{\alpha}\, (p_{12})^{\beta}
    + (p_{12})^{2} \, g^{\alpha\beta}_{[d]}
    \Big]\,,
\end{align}
in agreement with (2.13) of \cite{Gnendiger:2017pys}; the external
momentum is denoted by $p_{12}\!=\!p_1\!+\!p_2$. It is essential
that on the right hand side the quasi $d$-dimensional metric is
used. After the angular average, the resulting tensor can be
contracted with the rest of the diagram and the remaining phase-space
integrations can be performed. This procedure restores unitarity.
While it is fairly simple for our process, it is rather tedious to
implement for the general case.

In \HV\ the situation is very similar. In fact, the relations between
the partial results in \HV\ and \CDR\ are analogous to \eqref{eq:fdhvv}
and read
\begin{align}
  \label{eq:hvvscdr}
  \sigma^{(vv)}_{\mHV} &= 
  \frac{\sigma^{(vv)}_{\mCDR}}{1-\epsilon}\,, &
  \sigma^{(rr)}_{\mHV} &= 
  \frac{\sigma^{(rr)}_{\mCDR}}{1-\epsilon}\,, &
  \sigma^{(rv)}_{\mHV} &= 
  \frac{\sigma^{(rv)}_{\mCDR}}{1-\epsilon}\, . &
\end{align}
Again, combining the two-loop corrections and setting $d\!\to\!4$,
we obtain the same physical cross section $\sigma^{(2)}$ in \HV\ as in
the other schemes. However, the double-real term is plagued by precisely
the same unitarity-violating terms as in \FDH\ and therefore requires
the same unitarity-restoring treatment.

To summarize, \FDH\ and \HV\ can be used throughout for all parts of the
computation. The combination of the intermediate results then leads to the
same physical cross section. However, the treatment of the real corrections
in \FDH\ and \HV\ is much more involved than in~\DRED.\\

\section{Conclusions}
\label{sec:conclusions}

Extending previous studies to double-real and real-virtual
corrections, we have investigated the regularization-scheme
(in)dependence of a physical cross section at NNLO. As a case study we
have considered two-jet production in electron-positron annihilation
since for this process all computations can be done analytically.

One of the main findings is that in \DRED\ unitarity is preserved in
the sense that the double-real and real-virtual corrections are
obtained simply by integrating the corresponding four-dimensional real
matrix element squared over the $d$-dimensional phase space.  Terms of
$\mathcal{O}(\epsilon/\epsilon)$ that have been missed with respect to
\CDR\ are precisely compensated by corresponding changes in the
virtual contributions. As a consequence, the physical cross section is
scheme independent, as required.
This apparently magical conspiracy can be understood by systematically
disentangling the $\dim_s$-dimensional vector fields of \DRED\ into
$\dim$-dimensional gauge fields and $\Neps$-dimensional
$\epsilon$-scalar fields.  Like for any other fields, cross sections
involving $\epsilon$-scalars are finite and, as they come with a
multiplicity of $\Neps\!=\!2 \epsilon$, they vanish in the limit
$d\!\to\!4$.

Since the evanescent scalar fields do not contribute, an obvious way
forward is to not compute their contributions in the first place. This
corresponds to \CDR. Alternatively, their contributions can be
included. This corresponds to \DRED\ and, in fact, leads to
simplifications in some aspects of the computation. 
While it can be debated at length which scheme is simpler to apply,
the important main statement is that \DRED\ can be used systematically
and without undue complications for \textit{arbitrary} physical cross
sections at NNLO. The practical steps required for the computation of
the virtual, real, and real-virtual corrections are summarized as
follows:\\

\subsubsection*{Guideline for the evaluation of NNLO virtual
cross sections in DRED}

\begin{itemize}
 \item[(1)] Split gauge fields in the initial and final state as well as
  gauge fields that connect 1PI subdiagrams similar to \eqref{eq:split}
  into a quasi $\dim$-dimensional part and a quasi $\Neps$-dimensional part.
  As a result, there are diagrams without $\epsilon$-scalars (type~I) and
  diagrams where at least one gauge field is replaced by an $\epsilon$-scalar
  (type~II).
  
 \item[(2a)] Evaluate type I contributions by using the (quasi) four-dimensional
  \DRED\ algebra. Since there are no $\epsilon$-scalars at the tree-level,
  evanescent couplings have to be considered only at one-loop and their
  renormalization has at most to be known at one-loop as well.
  
 \item[(2b)] Evaluate type II contributions by using the IR divergence formula~%
 \eqref{eq:doubleVirtualEps}. The finite part follows from the fact that the
  tree-level cross section is proportional to at least one power of
  $\Neps\!=\!2\epsilon$. Similar to type I contributions, the renormalization
  of the evanescent couplings has at most to be known at one-loop order.
  In this way, no explicit two-loop computation with $\epsilon$-scalars has
  to be done.\\
\end{itemize}

\subsubsection*{Guideline for the evaluation of NNLO real
cross sections  in DRED}

\begin{itemize}
  \item[(1a)] Evaluate the real matrix element in $\dim_s\!=\!4$ dimensions
    which corresponds to setting $\epsilon\!\to\!0$ in the \CDR\ result.
    The subsequent phase-space integration is done in the usual way in
    $\dim$ dimensions.
  
  \item[(1b)] If the phase-space integration is split into a finite
    and a singular part (through subtraction or slicing), the integrand
    in the finite part is evaluated in four dimension in any
    scheme. However, in \DRED\ even in the singular part the matrix
    element is only needed in four dimensions. As a result, in
    \DRED\ the $\mathcal{O}(\epsilon)$ terms of the (double)
    unresolved limits (e.\,g.\ triple-collinear or double-soft) are
    not required.

\end{itemize}

\subsubsection*{Guideline for the evaluation of  NNLO real-virtual
  cross sections  in DRED} 

\begin{itemize}
  \item[(1)] Evaluate the real-virtual contributions by using the (quasi)
    four-dimensional \DRED\ algebra. The subsequent loop and phase-space
    integrations are done in $\dim$ dimensions.
  \item[(2)] Implement the UV renormalization by applying the split
    \eqref{eq:split} at the tree-level and by taking into account the
    one-loop renormalization of the (evanescent) couplings.
\end{itemize}

\noindent
Of course, it is also possible to use \FDH\ or \HV. The application of
these schemes to two-loop amplitudes is well understood and transition
rules between the various schemes are known. However, the naive use of
\FDH\ and \HV\ for real corrections at NNLO leads to unitarity violations.
In addition, the \FDH\ and \HV\ algebra gives rise to loop
momenta that are not $\dim$-dimensional, but actually strictly
four-dimensional. While it is possible to correct for these effects to
obtain consistent results, it leads to additional complications.

With ever more complicated final states to be tackled at NNLO, it might
well be worth to consider if computations in \DRED\ are not more
efficient, even in non-supersymmetric theories. In particular, the
presence of initial-state collinear singularities might further affect
these considerations. Their consistent treatment at NNLO in
\DRED\ remains to be worked out.

\section*{Acknowledgments} 
It is a pleasure to thank Dominik St{\"o}ckinger for insightful discussions.
We would like to thank Thomas Gehrmann for providing us with the form code
to compute the phase-space integration for the double-real corrections.

\bibliography{bibliography}{}
\bibliographystyle{JHEP}

\end{document}